\documentclass[12pt]{article}
\usepackage{amssymb}
\usepackage{epsf}
\usepackage{epsfig}
\usepackage{psfrag}
\usepackage{rotating}
\newcommand{\lsim}{
\mathrel{\hbox{\rlap{\hbox{\lower4pt\hbox{$\sim$}}}\hbox{$<$}}}}

\newcommand{\gsim}{
\mathrel{\hbox{\rlap{\hbox{\lower4pt\hbox{$\sim$}}}\hbox{$>$}}}}

\newcommand{\be}{\begin{equation}}
\newcommand{\ee}{\end{equation}}
\newcommand{\bi}{\begin{itemize}}
\newcommand{\ei}{\end{itemize}}

\newcommand{\IM}{{\rm Im}}

\begin{document}
\begin{titlepage}
\vspace*{-0.5truecm}

\begin{flushright}
TUM-HEP-572/04\\
hep-ph/0412195
\end{flushright}

\vspace*{0.3truecm}

\begin{center}
\boldmath

{\Large{\bf CP Asymmetry $A_{\psi K_S}$,  $\sin 2\beta$, the Sign of $\Delta
M_B$  
\vspace{0.3truecm}

and the Physics Beyond the Standard Model}}
\unboldmath
\end{center}

\vspace{0.4truecm}

\begin{center}
{\bf Monika Blanke and Andrzej J. Buras
} 
\vspace{0.4truecm}

 {\sl Physik Department, Technische Universit\"at M\"unchen,
D-85748 Garching, Germany}

\end{center}

\vspace{0.6cm}
\begin{abstract}
\vspace{0.2cm}\noindent
We demonstrate that the sign of the $B^0-\bar B^0$ mass difference $\Delta
M_B$ is irrelevant for the extraction of the angle 
$\beta=\phi_1$ of the unitarity triangle from the CP asymmetry 
$A_{\psi K_S}$. Only the weak phase in the mixing amplitude $M_{12}$ 
matters. Consequently the extraction of $\sin 2\beta$ done by BaBar and Belle
can only be affected by new weak phases in $M_{12}$ in the extensions of the 
SM and possibly but unlikely by new physics contributions to the decay
amplitude. 
\end{abstract}

\end{titlepage}

\thispagestyle{empty}
\vbox{}
\newpage

\setcounter{page}{1}
\pagenumbering{arabic}

\noindent
The CP asymmetry in $B_d^0\to\psi K_S$  
\be\label{ASYM}
A_{\psi K_S}(t)=\frac{\Gamma(\bar B^0_d(t)\to \psi K_S) -
\Gamma(B^0_d(t)\to \psi K_S)}
{\Gamma(\bar B^0_d(t)\to \psi K_S) +
\Gamma(B^0_d(t)\to \psi K_S)}
\ee
proposed by Bigi and Sanda \cite{BS} for the tests of CP violation 
in the $B$ meson system in 1980
has been observed \cite{BaBar,Belle}. Its value allows to extract
\be\label{sinb}
\sin 2\beta=0.726\pm 0.037~,
\ee
where $\beta=\phi_1$ is one of the angles of the unitarity triangle. This 
result is in an impressive agreement with the Standard Model (SM) expectations
\cite{BS,CKMCERN}.
Yet (\ref{sinb}) leads to a two fold ambiguity in the value of $\beta$ 
\be\label{bCKM}
\beta_{\rm CKM}=23.3\pm 1.6^\circ, \qquad 
\tilde\beta=\frac{\pi}{2}-\beta_{\rm CKM}
\ee
with the second possibility inconsistent with the SM expectations. Measuring
$\cos 2\beta$ will tell us which of these two solutions is chosen by nature.
The first direct experimental result of BaBar \cite{cos} for $\cos 2\beta$
and other analyses \cite{REF1,REF2,REF3} disfavour  
in fact the second solution in (\ref{bCKM}).

In extracting the value given in (\ref{sinb}) it has been assumed that the
mass difference $\Delta M_B=M_1-M_2 > 0$ with $M_1$ and $M_2$ denoting the
masses of the neutral $B$ meson eigenstates. As the
sign of $\Delta M_B$ is not known by itself, it is legitimate to ask 
what happens if $\Delta M_B$ was assumed to be negative. 

In a recent article Bigi and Sanda \cite{BS04} addressed the uniqueness
of the sign in (\ref{sinb}), reaching first the conclusion that it could 
not be determined irrespective of the sign of $\Delta M_B$ and  
that with $\Delta M_B<0$, the BaBar and Belle data imply 
$\sin 2\beta=-0.726\pm 0.037$. This would mean that in addition 
to (\ref{bCKM}) two 
additional solutions for the angle $\beta$ exist. These findings, 
if correct, would weaken significantly the present believe that the BaBar and
Belle data combined with the standard analysis of the unitarity triangle
imply that the CKM matrix \cite{CKM} 
is very likely the dominant source of CP violation
observed in low energy experiments.

In the first version of this note we challenged the conclusion of the authors 
of \cite{BS04} by
demonstrating that
\begin{itemize}
\item
The sign of $\Delta M_B$ is {\it irrelevant} for the determination of $\sin
2\beta$.
\item 
The only relevant quantity for this determination is the weak phase of the
mixing amplitude $M_{12}$. For a given phase convention of $B^0$ and 
$\bar B^0$, that cancels in the final expression for the asymmetry anyway,
the weak phase of $M_{12}$ is uniquely given in the SM 
and its possible extensions.
\end{itemize}

Meanwhile, the authors of \cite{BS04} have withdrawn their paper. 
Yet, the fact that even two well known experts in the field of CP violation 
had some doubts in this matter, shows that the issue of the sign of 
$\Delta M_B$ in the extraction of $\sin 2\beta$ is apparently not totally 
trivial. 
Therefore the following derivation should still be useful and moreover 
it will allow us
to summarize certain aspects of the sign of $\sin 2\beta$ beyond the SM that 
have been discussed at various places in the literature.

Using the expressions of \cite{AJBMFV,REV} 
but reversing the sign of $A_{\psi K_S}(t)$ 
in accord with (\ref{ASYM}) we have
\be\label{basic}
A_{\psi K_S}(t)=-\IM\xi\sin(\Delta M_B t),
\ee
where
\be
\xi=\frac{q}{p}\frac{A(\bar B^0\to f)}{A( B^0\to f)},
\qquad 
\frac{A(\bar B^0\to f)}{A(B^0\to f)}=-\eta_f e^{-i 2\phi_D}.
\ee
Here $\eta_f$ is the CP parity of the final state and $\phi_D$ is the 
weak phase in $A(B^0\to f)$, under the assumption that a single weak
phase dominates the decay amplitude. In the case at hand $\phi_D=0$ to a high
accuracy and $\eta_f=-1$. Consequently the second factor in $\xi$ can be set
to unity.

More importantly, we have
\be\label{E6}
\Delta M_B = M_1 - M_2 = 2 \mathrm{Re} \left( \frac{q}{p} M_{12} \right) = \pm
2|M_{12}|, 
\ee
\be\label{E7}
 B_1=pB^0+q\bar B^0, \qquad B_2 = pB^0 - q \bar B^0~, 
\ee
where $B_1$ and $B_2$ denote the mass eigenstates
and
\begin{equation}\label{q/p}
\frac{q}{p} = \pm \sqrt{\frac{M^*_{12}}{M_{12}}} = \pm \frac{M_{12}^*}{|M_{12}|}
=  2 \frac{M_{12}^*}{\Delta M_B}.
\end{equation}
Here we have used the fact that the width difference $\Delta \Gamma$ and 
$\Gamma_{12}$ can be neglected.
In  \cite{AJBMFV} the $\pm$ signs have been dropped. 
They are
unnecessary as will be seen in a moment but are kept here in view of 
the discussion initiated in \cite{BS04}.

Inserting (\ref{q/p}) into (\ref{basic}) we find
\begin{equation}\label{bas}
A_{\psi K_S}(t) = - 2 \mathrm{Im}\left(\frac{M^*_{12}}{\Delta M_B}\right)
\sin(\Delta M_B t).
\end{equation}
This formula demonstrates explicitly that 
the sign of $\Delta M_B$ is irrelevant and only the phase of 
$M_{12}$ matters.

Assuming that $M_{12}$ is governed by the usual $(V-A)\otimes(V-A)$ operator, 
we have
quite generally 
\begin{equation}\label{M12}
M_{12} = \frac{G_F^2}{12\pi^2}F_B^2
\hat B_B m_B M_W^2 (V_{td}^*V_{tb})^2 S_0(x_t)
\eta_B^{\mathrm{QCD}} r e^{i2\theta_d},
\end{equation}
where $F_B$ is the $B$ meson decay constant, $\hat B_B$ represents the 
hadronic matrix element of the $(V-A)\otimes(V-A)$ operator in question,
$S_0(x_t)>0$ the Inami-Lim function \cite{IL} and 
$\eta_B^\mathrm{QCD} \approx 0.55$ the QCD
correction \cite{BJW90,UKJS}. 
The last factor in (\ref{M12}) 
describes possible new physics contributions to the
Wilson coefficient of the $(V-A)\otimes (V-A)$ operator that have been 
discussed at
various occasions in the literature 
\cite{WPDELTAF2,Laplace,FLISMA,AI01,Berg,BSU}.
Without loss of generality we take $r>0$.
$\theta_d$ is a new weak phase.

Using
\begin{equation}
V_{tb}=1, \qquad V_{td}= |V_{td}|e^{-i\beta}
\end{equation}
and inserting (\ref{M12}) into (\ref{bas}) we find
\begin{equation}\label{master}
A_{\psi K_S}(t) = \mathrm{sign}(\hat B_B)~
\sin 2(\beta+\theta_d)~\mathrm{sign}(\Delta M_B) 
\sin(\Delta M_B t)~.
\end{equation}
This formula generalizes and summarizes various discussions of 
$A_{\psi K_S}(t)$ in
the SM and its simplest extensions that appeared in the 
literature. 
In particular in \cite{GKN}
 the relevance of the sign of $\hat B_B$ has been 
discussed. 
In these extensions only the
usual $(V-A)\otimes (V-A)$ operator is present and as new physics has no 
impact on its
matrix element between $B^0$ and $\bar B^0$ states, $\hat B_B>0$ 
\cite{CKMCERN}. 
 With
$\theta_d=0$ formula (\ref{master}) reduces to the usual formula used by 
BaBar and
Belle, except that sign$(\Delta M_B)$ in front of $\sin(\Delta M_B t)$ 
demonstrates
that the sign of $\Delta M_B$ is immaterial.

With $\theta_d=90^\circ$ one recovers a particular minimal flavour violation 
scenario of 
\cite{BF01} in
which the sign of $S_0(x_t)$ is reversed. In this case indeed the BaBar and 
Belle
measurement implies $\sin 2 \beta = -0.726\pm0.037$, but this has nothing to do
with the sign of $\Delta M_B$.

\vspace{1cm}

\noindent
{\bf Acknowledgements}\\
\noindent

A.J.B would like to thank Yossi Nir for a discussion and 
stressing the importance
of the clarification of the claim made by the authors of \cite{BS04}.  

This work has been supported by
Bundesministerium f\"ur
Bildung und Forschung under the contract 05HT4WOA/3  and the 
GIF project G-698-22.7/2002.

\end{document}